\documentclass[letterpaper,twocolumn,prl,aps,superscriptaddress,floatfix]{revtex4}
\usepackage[latin1]{inputenc}
\usepackage{bm}
\usepackage{amssymb,amsbsy,amsmath}
\usepackage{graphicx}
\usepackage{verbatim}
\usepackage{color}
\usepackage{pifont}
\usepackage{hyperref}
\usepackage{xcolor}

\begin{document}
	
\title{Inverse Weak measurement in SERF magnetometer}
	
\author{Qian Cao}
\affiliation{ National Institute of Extremely-Weak Magnetic Field Infrastructure, Hangzhou, 310051, China}

\author{Liang Xu}
\email{liangxu.ceas@nju.edu.cn}
\affiliation{National Laboratory of Solid State Microstructures, Key Laboratory of Intelligent Optical Sensing and Manipulation, College of Engineering and Applied Sciences, and Collaborative Innovation Center of Advanced Microstructures, Nanjing University, Nanjing 210093, China}
	
\author{Ziqian Yue}
\affiliation{Zhejiang Provincial Key Laboratory of Ultra-Weak Magnetic-Field Space and Applied Technology, Hangzhou Innovation Institute, Beihang University, Hangzhou, 310051, China}
\affiliation{Key Laboratory of Ultra-Weak Magnetic Field Measurement Technology, Ministry of Education, the School of Instrumentation and Optoelectronic Engineering, Beihang University, Beijing 100191, China}

\author{Jianqi Yang}
\affiliation{Zhejiang Provincial Key Laboratory of Ultra-Weak Magnetic-Field Space and Applied Technology, Hangzhou Innovation Institute, Beihang University, Hangzhou, 310051, China}
\affiliation{Key Laboratory of Ultra-Weak Magnetic Field Measurement Technology, Ministry of Education, the School of Instrumentation and Optoelectronic Engineering, Beihang University, Beijing 100191, China}

\author{Yueyang Zhai}
\email{yueyangzhai@buaa.edu.cn}
\affiliation{Zhejiang Provincial Key Laboratory of Ultra-Weak Magnetic-Field Space and Applied Technology, Hangzhou Innovation Institute, Beihang University, Hangzhou, 310051, China}
\affiliation{Key Laboratory of Ultra-Weak Magnetic Field Measurement Technology, Ministry of Education, the School of Instrumentation and Optoelectronic Engineering, Beihang University, Beijing 100191, China}
\affiliation{Hefei National Laboratory, Hefei 230088, China}

\date{\today}
\maketitle
	
{\bf{Weak measurement techniques have been extensively applied in the field of quantum precision measurement to detect ultra-small signals due to the amplification effect. In this work, we propose an optical detection system for a spin-exchange relaxation-free (SERF) magnetometer based on the inverse weak measurement (IWM) framework. By using the spatial pattern of a probe laser as the measurement pointer, we successfully detect ultra-weak magnetic fields. In our model, the spatial pattern of the probe laser is weakly coupled to its polarization, which is sensitive to external magnetic fields. Through post-selection on the optical polarization, the ultra-small magnetic field is significantly amplified with the amplification factor inversely proportional to the coupling strength, as reflected in the measured displacement of the final spatial pattern. By analysing the response curve of the probe laser displacement to the magnetic field, we identify the point of maximum sensitivity, achieving a magnetic field sensitivity of 182.8 $fT/Hz^{1/2}$. Furthermore, in the IWM scheme, the detected signals depend only on the internal degrees of freedom of the probe laser, making the system robust against the fluctuations in laser power. To demonstrate this advantage, we compute the Allan standard deviation of the output signals for both conventional and IWM detection methods. The results indicate that the IWM-based method improves stability of detection by one to two orders of magnitude. This work presents a novel detection approach that integrates weak measurement techniques, offering a significant enhancement in the performance of SERF magnetometers.
}}
	
\section{I. Introduction}
The spin-exchange relaxation-free (SERF) magnetometer is a kind of ultra-sensitive magnetic field sensor that plays a crucial role in the field of quantum precision measurement~\cite{1-07 Nature}. With its exceptional capability to detect ultra-weak magnetic fields, the SERF magnetometer demonstrates significant potential in fundamental scientific research~\cite{1-2 25xuprl}, which is helpful for the dark matter detection~\cite{2-24 JiangM Dark,3-21 Su Dark} and material structure analysis~\cite{4-24 Sun IEEE}. In addition, with the improvement of extremely weak magnetic field detection technology, the SERF magnetometer is further reduced and integrated in volume, which can be utlized into biomagnetic field measurement~\cite{5-24 Y Heart} and geomagnetic navigation~\cite{6-24 Xu Earth}. It has been successfully used for early diagnosis and prevention of cardiovascular and cerebrovascular diseases~\cite{6-24 heart,6-24 breath}. Thus, the SERF magnetometer is a hot research point in quantum precision measurement region.

In the probe laser system of a SERF magnetometer, the rotation angle of a linearly polarized probe laser is typically used to detect the ultra-small magnetic field~\cite{7-13 ShengD,8-67 Happer}. Based on the theory of optical rotation angle detection, various methods, such as Faraday modulation detection~\cite{9-2014 F Faraday}, magnetic field modulation detection~\cite{10-2018 Jiang Magnetic}, photoelastic modulation detection~\cite{11-2007 Romalis,12-2023 Fang1} have been widely applied in the design of SERF magnetometer probe laser systems. Additionally, probe laser interference~\cite{13-2023 Feedback} and multi-reflection cavitiese~\cite{13-2025 multic} are demonstrated to be effective to enhancing the optical rotation angle signal. In these methods, the optical power of the probe laser, after passing through the vapor cell, is used to measure the optical rotation angle, which corresponds to the strength of the detected magnetic field~\cite{14-2024 Fang2}. However, slow drifts in the optical power of the probe laser introduce corresponding drifts in the detected signal, thereby limiting the signal stability of the SERF magnetometer~\cite{15-2021 X Noise}. Consequently, there is a critical need to develop detection methods that mitigate the influence of slow optical power drifts and improve the stability of SERF magnetometer signals.

Weak measurement (WM) has been experimentally demonstrated to be superior in suppressing technical noise~\cite{16-2014 Weakadvantages,16-2020 XuliangPRL}. Since WM theory was first proposed by Aharonov, Albert based on the Von Neumann measurement theory~\cite{17-1988 Weaktheory}, it has been widely applied in the field of optical precision measurements, including angle~\cite{18-2021 Weak-angle}, displacement~\cite{19-2003 Weak-displacement}, phase measurement~\cite{20-2013 Weak-phase1,21-2010 Weak-phase2}, radio detection~\cite{22-2023 Weak-Radion}, among others. Selecting an appropriate measurement pointer based on the characteristic of the measured parameter is crucial~\cite{23-2024 Weak-Progess}. For instance, Yin et al utilized the broadband light source as a measurement pointer and successfully achieved magnetic field measurements~\cite{24-2021 YinP}. However, for the narrow linewidth characteristics of the probe laser in SERF atomic magnetometers, spectrum-based measurements are challenging to implement and the sensitivity in these methods has not reached a higher level~\cite{24-2025 Zhangzhiyou}. On the other hand, the spatial pattern of a laser beam remains unaffected by fluctuations in optical power~\cite{25-2024 HengX}. Therefore, using the spatial pattern as a measurement pointer for magnetic field weak measurement holds promise for improving the stability of magnetic field signals, thus showing significant potentials in SERF atomic magnetometers~\cite{26-2024 arxiv}.

In this manuscript, a new method for detecting magnetic fields in SERF atomic magnetometers based on the inverse weak measurement (IWM) theory is proposed. The laser spatial pattern is employed as the measurement pointer to facilitate the detection of magnetic fields in the SERF atomic magnetometer. In Section II, the model of the IWM-enhanced SERF magnetometer is established. Based on this model, the optical implementation is designed in Section III. Finally, in Section IV, it is demonstrated that magnetic fields can be detected by measuring the displacement of the probe laser. Using this approach, we achieved a sensitivity of $182.8 fT/ Hz^{1/2}$ with excellent stability against laser power fluctuations. These findings provide valuable insights for advancing the study and development of probe laser systems in SERF atomic magnetometers.

\begin{figure*}[hbt]
\centering
\includegraphics[width=0.98\textwidth]{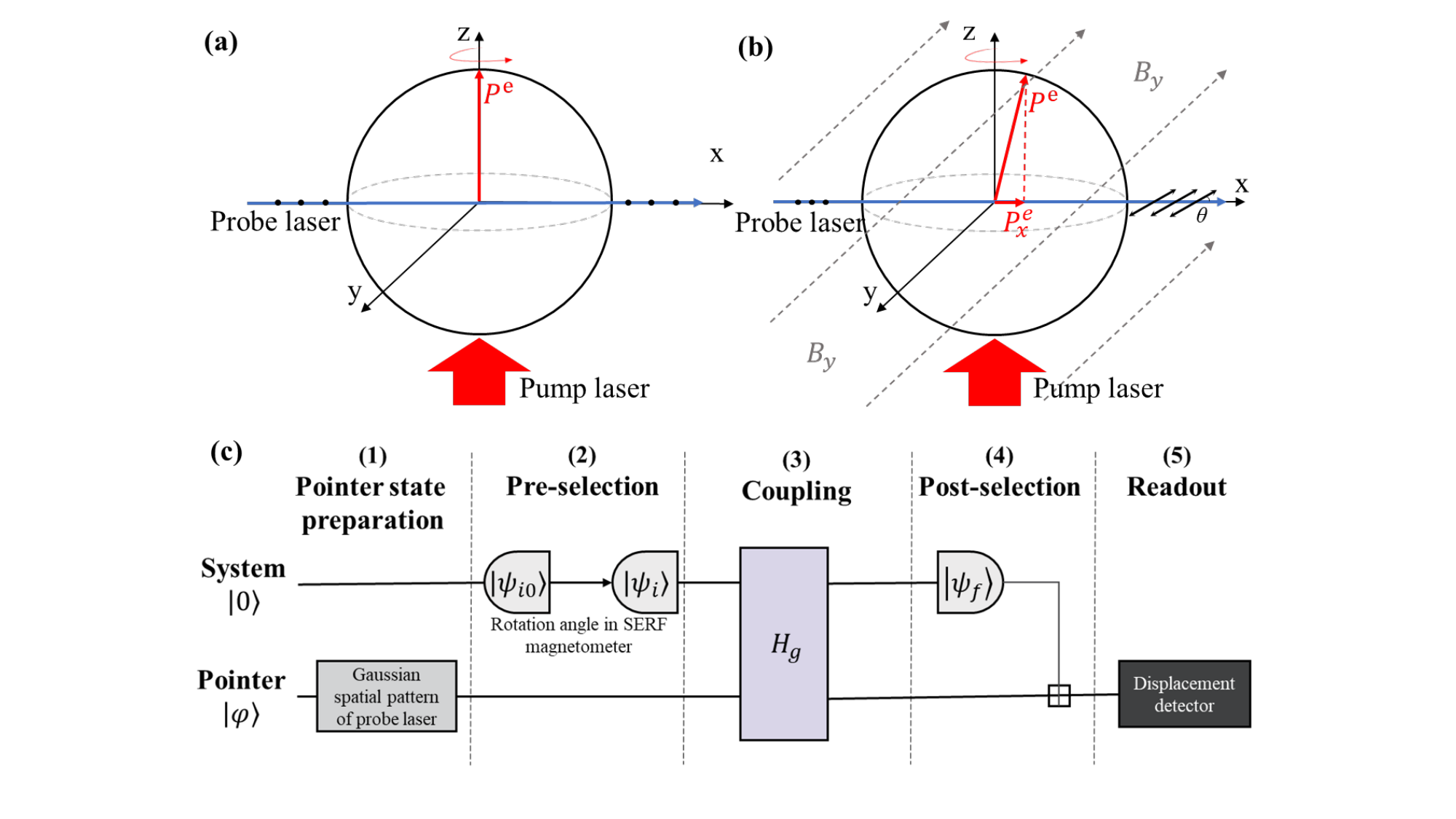}
\caption{(a) The spin process of alkali metal atoms which is pumped by pump laser; (b) The spin process of alkali metal atoms in external magnetic field $B_y$; (c) The logic diagram of Inverse Weak measurement in SERF magnetometer. (As shown in Fig.\ref{Fig1}(a), once the alkali metal atoms in high temperature are pumped by a beam of pump laser, and the spin process of them is in same direction along the pumping direction. At the same time, the polarization state of probe laser will not be changed after passing through the alkali metal atoms. As shown in Fig.\ref{Fig1}(b), when an external magnetic field is added along the y-axis exists, an offset of electron spin polarization vector $P^e$ from pumping direction is produced and a projection $P_{x}^e$ is generated in X-axis. The value of $P_{x}^e$ is related to the added magnetic field $B_y$. Based on the Eq.\eqref{eq3} and the theory listed, the polarization plane of linearly polarized probe laser is rotated and the rotated angle $\theta$ can be expressed to the equation including $P_{x}^e$.)
}
\label{Fig1}
\end{figure*}
	
\section{II. Theory model}
\textbf{Magnetic field measurement theory in SERF magnetometer.} In a SERF magnetometer, the magnetic field is detected through its interaction with alkali metal atoms in the SERF state, which causes the linearly polarized probe laser to rotate in this polarization plane. The rotation angle corresponds to the magnetic field being measured. The relationship between the detected magnetic field and the alkali metal atoms is determined by the Bloch's equation
\begin{equation}
\frac{d}{{dt}}P = \frac{1}{Q}[{\gamma _{\rm{e}}}B \times P + {R_{{\rm{OP}}}}(\frac{1}{2}s-P) - {R_{{\rm{rel}}}}P],
\label{eq1}
\end{equation}
where, $P$ is the spin polarization vector of alkali metal electrons; $\gamma_e$ is the gyromagnetic ratio of the alkali metal electrons; $q$ is the nuclear slowing-down factor and is influenced by the electron spin polarization vector; $B$ is magnetic field vector at alkali metal atoms; $R_{OP}$ is optical pumping rate and $R_{rel}$ is atomic spin relaxation rate; $s$ is average photon spin.
	
By solving the steady-state solution of Bloch's equation, the relationship between the detected magnetic field and alkali metal atoms can be established. The spin polarization vector of electrons along the probe direction can be expressed as:
\begin{equation}
P_x^e = {P_0}\frac{{{\gamma _{\rm{e}}}}}{{{R_{{\rm{OP}}}} + R{}_{{\rm{rel}}}}}{B_{\rm{y}}},
\label{eq2}
\end{equation}
where $P_x^e$ is the spin polarization vector of alkali metal electrons in X-axis; $P_0$ is the initial value of the atomic spin vector; $B_y$ is the magnetic field along the Y-axis, which is the sensitive axis in this study. Equation~\eqref{eq2} shows that the spin polarization vector is linearly related to the detected magnetic field. This demonstrates that the magnetic field influences the spin precession process of the alkali metal atoms, providing the theoretical foundation for magnetic field detection in SERF magnetometers.
	
\begin{figure*}[hbt]
\centering
\includegraphics[width=0.98\textwidth]{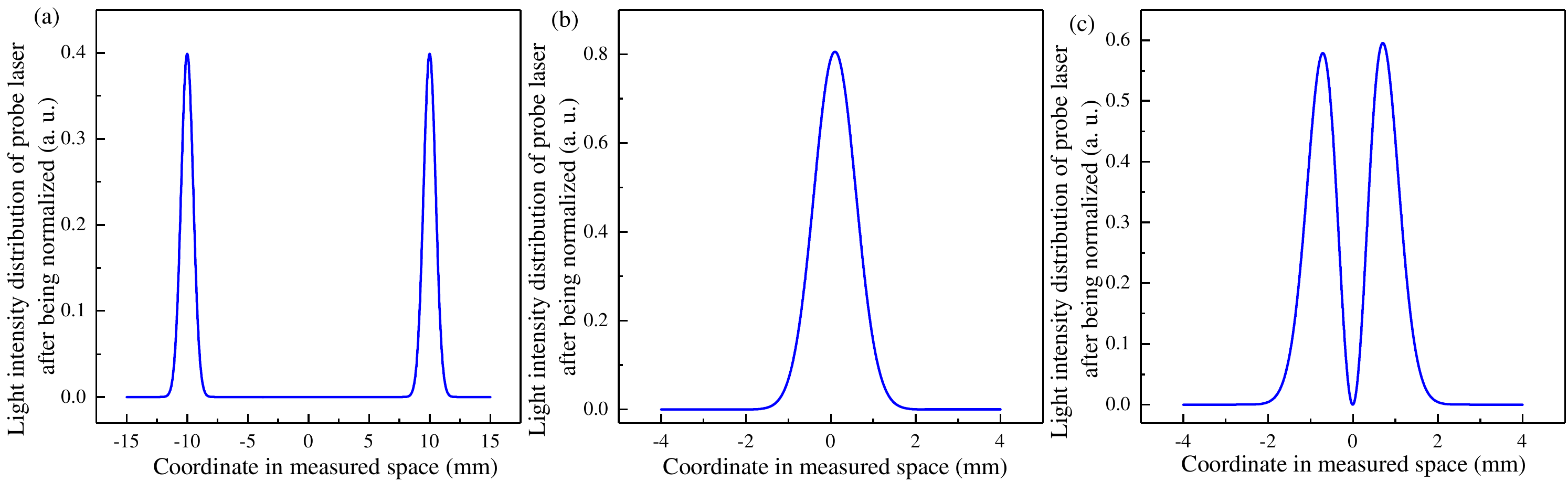}
\caption{The spatial pattern of the probe laser after post-selection. (a) Strong measurement regime: the displacement is directly proportional to the coupling strength $g$. (b) Weak-value amplification (WVA) regime: the coupling strength $g$ is effectively amplified by the weak value, leading to an enhanced displacement. (c) Inverse weak measurement (IWM) regime: the displacement of the final meter state (MS) is inversely proportional to both the weak value and the coupling strength. In this regime, the parameter encoded in the weak value is amplified by using a small coupling strength $g$. The post-selected MS exhibits a bi-peak structure, where changes in the parameter alter the relative height of the peaks. The parameter can be extracted by measuring the mean displacement of the final MS.}
\label{Fig2}
\end{figure*}
	
As shown in Fig.\ref{Fig1} (a) and (b), when a linearly polarized probe laser passes through the alkali metal atoms, a rotation angle $\theta$ of probe laser is produced. This process can be described using the theory that a linearly polarized laser rotates after passing through a birefringent medium~\cite{27-2024 angle probe}. The resulting rotation angle is given by 
\begin{equation}
\theta  = \frac{\pi}{2}l{r_e}c{n_{alkali}}P_x^e f_{D1}\frac{\left(\upsilon_{probe}-\upsilon_{D1}\right)}{\left(\upsilon_{probe}-\upsilon_{D1}\right)^2+\left(\Gamma_{D1}/2\right)^2},
\label{eq3}
\end{equation}
which establishes the relationship between the probe laser's rotation angle and the detected magnetic field via the electron spin polarization vector along the probe direction. Here, $l$ represents the length of the cell; $r_e$ is the classical electron radius; $c$ is the speed of light; $n_{alkali}$ is the density of alkali metal atoms; The factor $f_{D1}=\frac{2}{3}$; $\upsilon_{probe}$ corresponds to the frequency of the probe laser. $\upsilon_{D1}$ is the resonance frequency of the first line of Rb atoms, and $\Gamma_{D1}$ is the pressure broadening of the alkali metal atoms. Therefore, magnetic field can be detected by measuring the rotation angle of probe laser actually.
	
In conventional detection methods, the polarization angle of the probe laser is measured by passing the laser through a polarizer. The polarizer transforms the polarization angle information to an intensity difference, rendering the measured signal directly proportional to the optical power of the output probe laser. Consequently, fluctuations or drifts in the optical power of the probe laser can significantly impact the accuracy and stability of the measured polarization angle. This intrinsic dependence on optical power introduces potential errors and undermines the robustness of the measurement system, particularly in applications demanding high precision and long-term stability~\cite{28-2019 angle probe}.
	
\textbf{Inverse weak measurement theory in SERF magnetometer.} 
In this manuscript, we present a novel approach to magnetic field measurement utilizing the displacement of a probe laser, grounded in the theory of IWM. As depicted in Fig. \ref{Fig1} (c), the polarization degree of freedom of photons is treated as the quantum system (QS), which captures signals from the SERF magnetometer. Meanwhile, the Gaussian spatial profile of the probe laser serves as the meter state (MS), described as
\begin{equation}
|\Phi(q) \rangle  =  \int dq\frac{1}{{{{\left( {2\pi {\sigma ^2}} \right)}^{1/4}}}}{\rm{exp}}\left( { - \frac{{{q^2}}}{{4{\sigma ^2}}}} \right)\left| q \right\rangle,
\label{eq4}
\end{equation}
where $\sigma$ denotes the standard deviation of the Gaussian beam, and $q$ represents the spatial position in the measurement space. The initial QS state is prepared as $\left| {{\psi _{i0}}} \right\rangle  = \frac{1}{{\sqrt 2 }}\left( {\left| H \right\rangle  + \left| V \right\rangle } \right)$, where $\left| H \right\rangle$ and $\left| V \right\rangle$ denote horizontal and vertical polarization states, respectively. Upon interaction with the alkali metal atoms in the SERF magnetometer, the initial state evolves into the pre-selected state:
\begin{equation}
\left| {{\psi _i}} \right\rangle  = \rm{cos}\left( {\pi /4 - \theta } \right)\left| H \right\rangle  + sin\left( {\pi /4 - \theta } \right)\left| V \right\rangle,
\label{eq5}
\end{equation}
where $\theta$ is a parameter related to the magnetic filed and is defined in Eq. \eqref{eq3}. The QS then interacts with the MS via a von Neumann-type interaction Hamiltonian $\hat{H} = g \delta(t-t_0)\hat{A}\otimes \hat{P}$ with the coupling strength $g$. Here, $\hat{A} = |H\rangle\langle H| - |V\rangle\langle V|$ is the QS observable, and $\hat{P}$ is the momentum operator of the MS. This interaction induces a unitary evolution $\hat{U}_{jt} = \exp(-i\int \hat{H} dt) = \exp(-ig\hat{A}\otimes \hat{P})$, resulting in the joint state of the QS and the MS $|\Phi_{jt}\rangle = \hat{U}_{jt} |\psi_i\rangle |\Phi\rangle$. Subsequent post-selection of the QS into the final state $|\psi_f\rangle = (|H\rangle - |V\rangle)/\sqrt{2}$ yields the final MS $|\Phi_f\rangle = \langle \psi_f|\Phi_{jt}\rangle/\sqrt{p_f}$, where $p_f$ is the success probability of post-selection. In this framework, the weak value of the QS observable $\hat{A}$ is defined as
\begin{equation}
\langle \hat{A}\rangle_w = \frac{\langle \psi_f \hat{A}|\psi_i\rangle}{\langle \psi_f|\psi_i\rangle} = \cot \theta.
\label{eq6}
\end{equation}
By measuring the final MS in the $q$-space, we derive the probability distribution $P_q = |\langle q|\Phi_f\rangle|^2$. From this, the mean displacement of the MS is given by
\begin{equation}
\Delta Q = \int q P_q dq = \frac{g\sin 2\theta}{1 - \exp(-\frac{g^2}{2\sigma^2})\cos 2\theta}.
\label{eq7}
\end{equation}

The relationship between the coupling strength $g$, the initial MS fluctuation $\sigma$, and the weak value $\langle \hat{A}\rangle_w$ allows us to approximate $\Delta Q$ in three distinct regimes:
\begin{equation}
\Delta Q  \approx 
\begin{cases} 
g\sin2\theta , & \text{if } g \gg \sigma \\
g \cot \theta , & \text{if } g \ll g \langle \hat{A}\rangle_w \ll \sigma \\
\frac{4\sigma^2}{g}\theta , & \text{if } g \ll \sigma \ll g \langle \hat{A}\rangle_w
\end{cases},
\label{eq8}
\end{equation}
which correspond to the strong-measurement, weak-value amplification (WVA) and IWM regimes, respectively. 
	
In our protocol, the parameter $\theta$ which encodes information about the magnetic field is amplified using the IWM regime by setting a small coupling strength $g$. The spatial patterns of the post-selected MS under the three regimes are illustrated in Fig. \ref{Fig2}. Instead of relying on conventional power detection schemes, this approach infers $\theta$ from the displacement of the MS. Consequently, this method is robust against both short-term and long-term laser power fluctuations. The IWM scheme effectively transforms power-based detection into position-based detection for measuring the small phase shifts induced by the SERF magnetometer. This amplification mechanism significantly reduces the sensitivity requirements of the detector, enabling ultra-high sensitivity in magnetic field measurement. The proposed method holds great potential for applications in precision magnetometry and quantum sensing technologies.

\section{III. Experiment}
	
\begin{figure*}[hbt]
\centering
\includegraphics[width=1\textwidth]{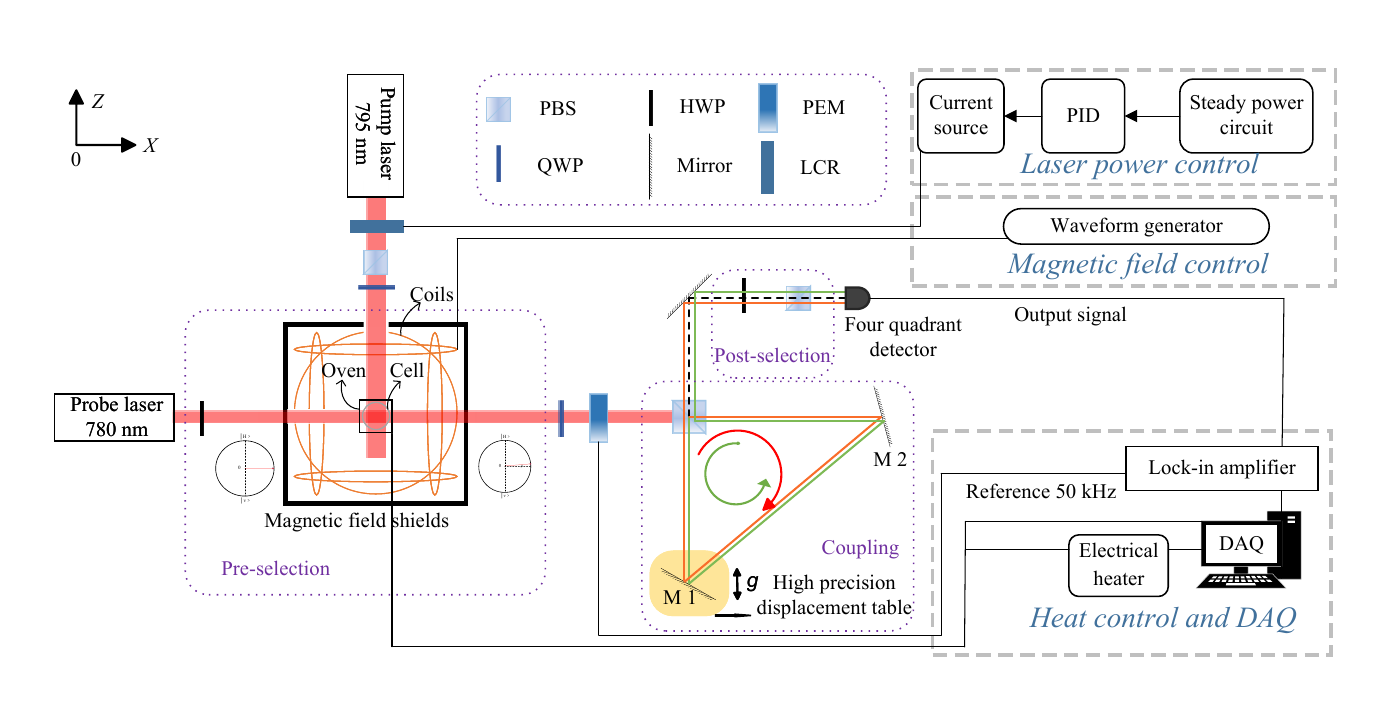}
\caption{Experimental setup. The SERF SERF magnetometer based on the IWM can be divided into three sections, enclosed by the purple dashed boxes, representing the three stages of weak measurement: pre-selection, coupling, and post-selection. The alkali metal gas chamber is positioned at the intersection of the probe and pump laser beams. Surrounding the gas chamber are the magnetic field compensation coils (orange coils) and magnetic field shields (black squares). PBS: polarizing beam splitter; HWP: half-wave plate; QWP: quarter-wave plate; M1 and M2: mirrors in the Sagnac interference loop; LCR: liquid crystal half-wave variable retarder; PEM: photoelastic modulator. 
}
\label{Fig3}
\end{figure*}

The experimental setup of the SERF magnetometer based on the IWM is illustrated in Fig.\ref{Fig3}. At the core of the SERF magnetometer platform, a rubidium (Rb) vapor cell serves as the sensing element for the detected magnetic field. To keep the Rb atoms in the SERF state, both high temperature and an ultra-low magnetic field are essential. The temperature of the Rb cell is precisely stabilized at 150 $^\circ$C using a boron nitride (BN) oven paired with a custom-designed heater system. An ultra-low magnetic field environment is achieved through a five-layer $\mu$-metal magnetic field shield combined with magnetic field compensation coils, ensuring the magnetic field at the cell center is maintained at the nanotesla (nT) level. Especially, the configurations of magnetic field compensation coils are designed as Lee-Whiting coils and Saddle coils, which is controled by the high precision current source. Once the Rb atoms are stabilized under these conditions, a circularly polarized pump laser beam is directed into the Rb cell along the Z-axis. The pump laser is tuned to the Rb D1 resonance line (794.98 nm) to align the atomic spins along the pumping direction, as illustrated in Figs. \ref{Fig1}(a) and \ref{Fig1}(b). To minimize the impact of the alkali-metal atomic pumping effect on magnetic field detection, the pump laser power is stabilized using a liquid crystal retarder (LCR) and a custom-built laser power control system, as depicted in Fig. \ref{Fig3}.

As described in Section II, the coupling strength $g$ and the rotation angle $\theta$ of the probe laser induced by the magnetic field result in changes to the spatial pattern of the probe laser, specifically manifesting as variations in laser displacement.The probe laser, generated by a distributed Bragg reflector laser, is horizontally polarized with a Gaussian spatial profile described by Eq. \eqref{eq4}. A half-wave plate (HWP) rotates the polarization angle to 45$^\circ$, preparing the linearly polarized probe beam in the initial state $\left| \psi_{i0} \right\rangle$. After interacting with the Rb atoms in the SERF state, the probe beam evolves into the pre-selected state described by Eq. \eqref{eq5}. Then, the probe beam enters the Sagnac interferometer. With the Sagnac loop, the $\left| H \right\rangle$ component is transmitted through the polarizing beam splitter (PBS) in the clockwise direction, while the $\left| V \right\rangle$ component is reflected in the counterclockwise direction. The two beams are subsequently recombined at the PBS. By introducing a small displacement $\sqrt{2}g$ to the mirror 1 mounted on a sub-micron precision translation stage inside the Sagnac interferometer, the horizontal and vertical components are separated by $2g$ to implement the weak-measurement Hamiltonian. The post-selection is realized through a combination of a HWP at $67.5^\circ$ and a PBS. Finally, a quadrant photodetector is positioned at the dark port to measure the laser displacement. In order to suppress low-frequency noise, a PEM is utlized into the probe laser system shown in Fig.\ref{Fig3}, which can add a 50 kHz modulated reference signal. Through demodulating the micro displacement information of output signal in lock-in amplifier, the detected magnetic field is calculated and recorded. 

\begin{figure*}[hbt]
\centering
\includegraphics[width=0.8\textwidth]{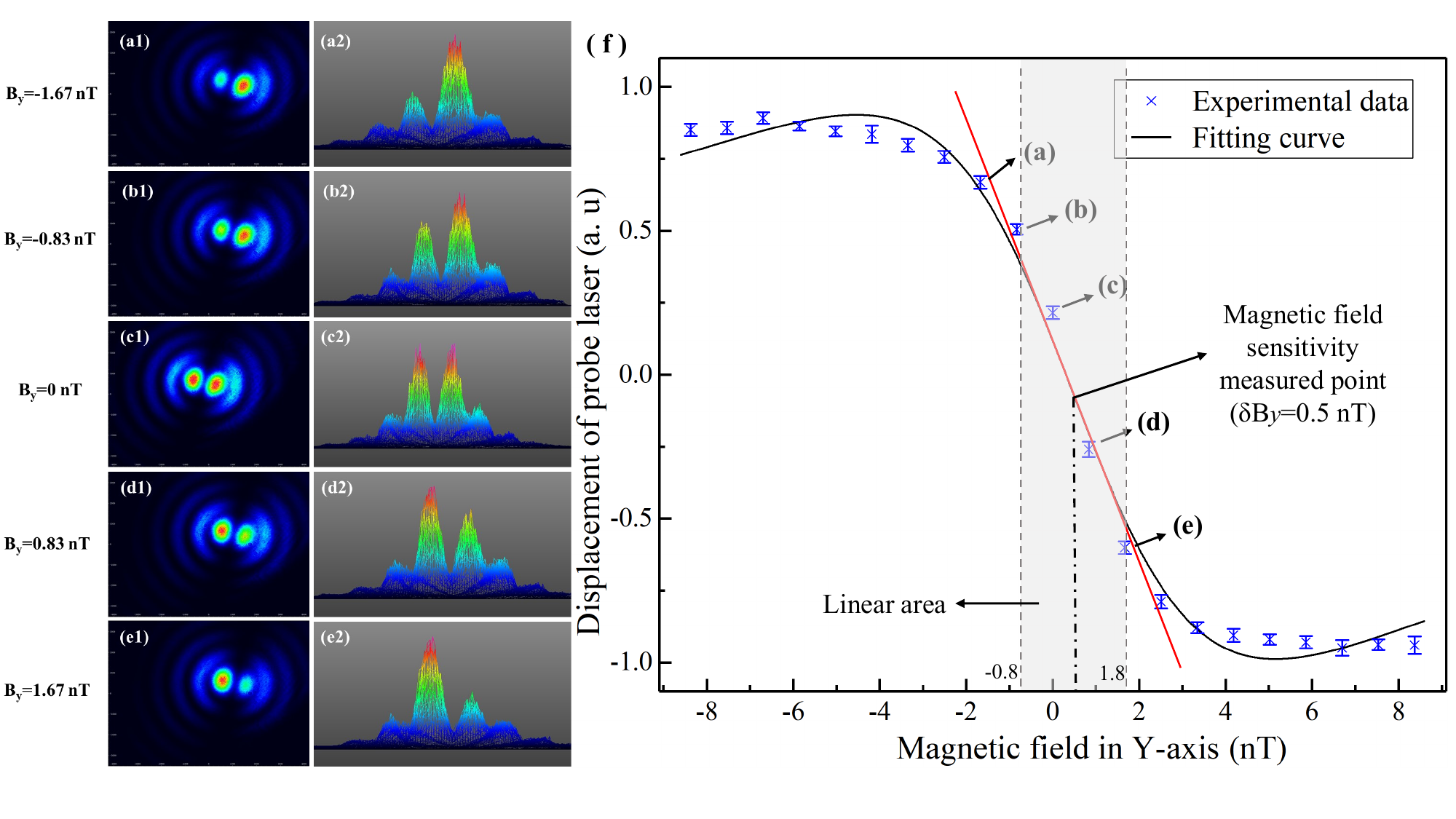}
\caption{(a)-(e) Laser spatial patterns of the output laser in the dark port under different magnetic field strengths. (f) Displacement of the output laser in the dark port at various magnetic field points. The horizontal axis represents the magnetic field strength deviating from the magnetic compensation point, which is provided by magnetic field compensation coils denoted as $\delta B_{y}$. The vertical axis represents the normalized displacement of output laser.}
\label{Fig4}
\end{figure*}
	
\section{IV. Results and Discussion}
 
After completing the setup of the SERF magnetometer as described in Section III, including adjustments such as heating temperature and residual magnetic field compensation process, the compensated magnetic fields along the three axis were recorded as follows: $B_{x}=1.11$ nT, $B_{y}=0.49$ nT, $B_{z}=0.86$ nT. These values correspond to the magnetic field compensation points, where the magnetic field at the center of the system is approximately zero. Subsequently, the M1 mirror and the 5-degree-of-freedom displacement table were adjusted to minimize the optical power of the output laser in the dark port. At this point, the M1 mirror was further fine-tuned using a precision displacement table to achieve a bimodal pattern in the output laser's dark port, as shown in Fig. \ref{Fig4} (c).

\begin{figure}[ht]
\centering
\includegraphics[width=0.47\textwidth]{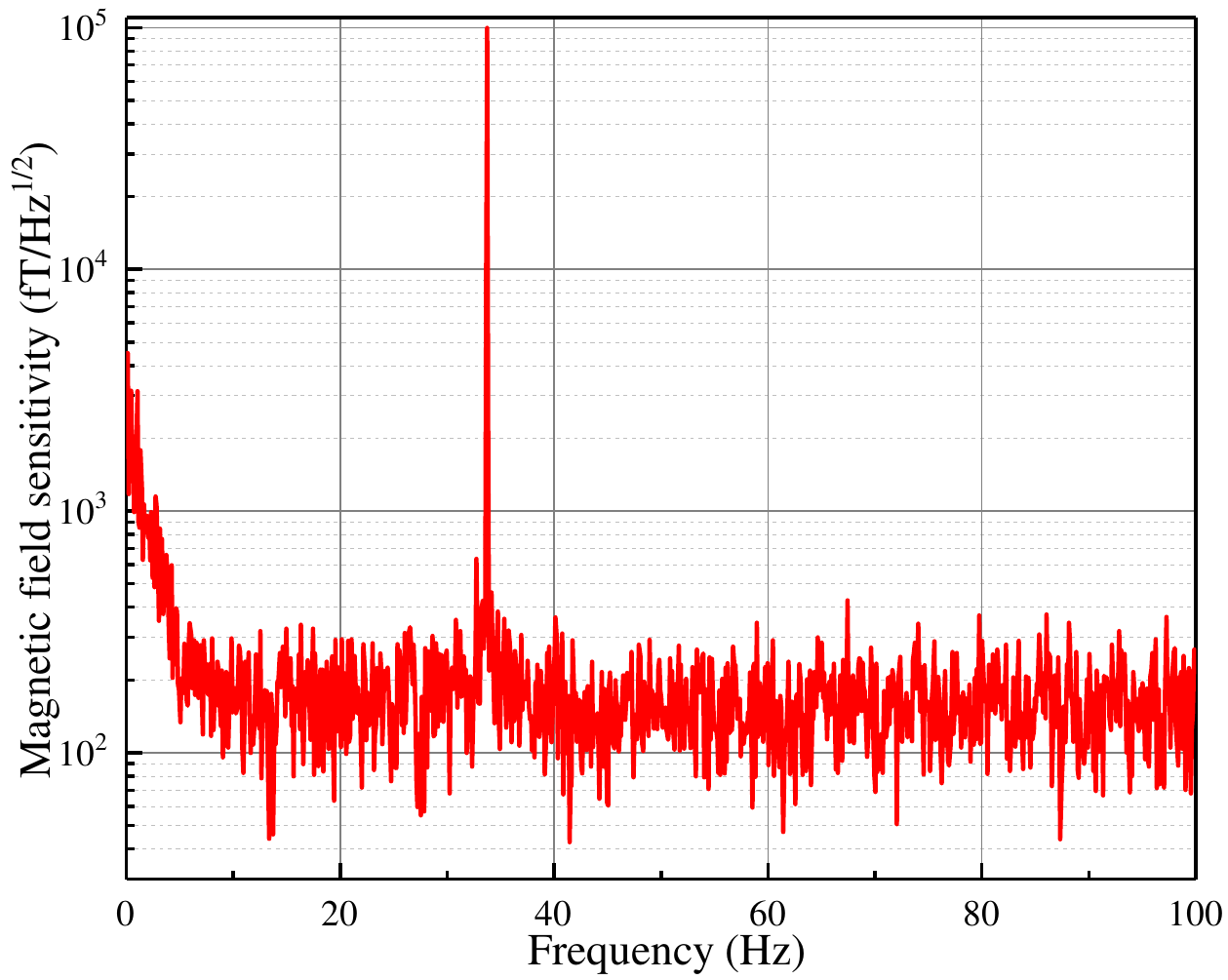}
\caption{Sensitivity of the IWM-enhanced magnetometer. The output signal, representing the displacement of the probe laser, is collected by a four-quadrant detector. The raw signal, initially in the form of a current, is converted into a magnetic field signal based on its response to a reference magnetic field.}
\label{Fig5}
\end{figure}

As illustrated in Fig.\ref{Fig4} (a)-(e), five magnetic field points ($\delta B_{y}=-1.67, -0.83, 0, 0.83, 1.67$ nT) were selected, and the spatial pattern of the output laser in the dark port at these points are captured using a beam quality analyzer. Due to diffraction effects, fluctuation peaks appear at the edges of the laser spatial pattern, but these do not affect the extraction of the central peak characteristics. At $\delta B_{y}=0$ nT, the two peaks of the probe laser in the dark port are nearly symmetrical. As the magnetic field along the Y-axis increases, the difference between the two peaks grows, following a consistent trend. Moreover, the direction of the detected magnetic field determines whether the first peak is higher than the second, or vice versa. Different spatial patterns of the probe laser correspond to distinct displacements and directions, as quantified in Fig. \ref{Fig4}(f). 

Using Eq. \eqref{eq8}, the displacement is found to be linearly related to the polarization rotation angle of the probe laser, resulting in a dispersion curve that matches the experimental results obtained from both the traditional detection method and the IWM. Fig. \ref{Fig4}(f) further demonstrates the sensitivity of displacement to magnetic field variations. Fitting the curve reveals that the slope of displacement versus magnetic field is maximized at approximately 0.5 nT. Due to system shifts, such as those caused by the current amplification process, the point of maximum slope does not coincide with the magnetic field compensation point. This point, referred to as the "sensitivity detection point" in this study, exhibits a 0.5 nT shift from the compensation point. By applying a magnetic field along the Y-axis and measuring the sensitivity at the sensitivity detection point, the magnetic field sensitivity is determined to be 182.8 $fT/Hz^{1/2}$, as shown in Fig. \ref{Fig5}. Furthermore, the linear operating range, spanning from -0.8 nT to 1.8 nT, is highlighted in gray in Fig. \ref{Fig4}(f).

The sensitivity measurement method employed here is consistent with that used in traditional SERF magnetometers~\cite{29-2024 angle probe}. A 30 Hz, 100 nT sinusoidal magnetic field was applied along the Y-axis using the Y-compensation coils. The lock-in amplifier demodulated the output signal, which was subsequently subjected to power spectral density analysis. The frequency response results are presented in Fig. \ref{Fig5}. It was experimentally confirmed that magnetic fields induce changes in the laser's spatial patterns, demonstrating the feasibility of using the IWM for magnetic field measurements. This approach lays the foundation for designing an IWM-enhanced magnetometer.
	
\begin{figure}[ht]
\centering
\includegraphics[width=0.47\textwidth]{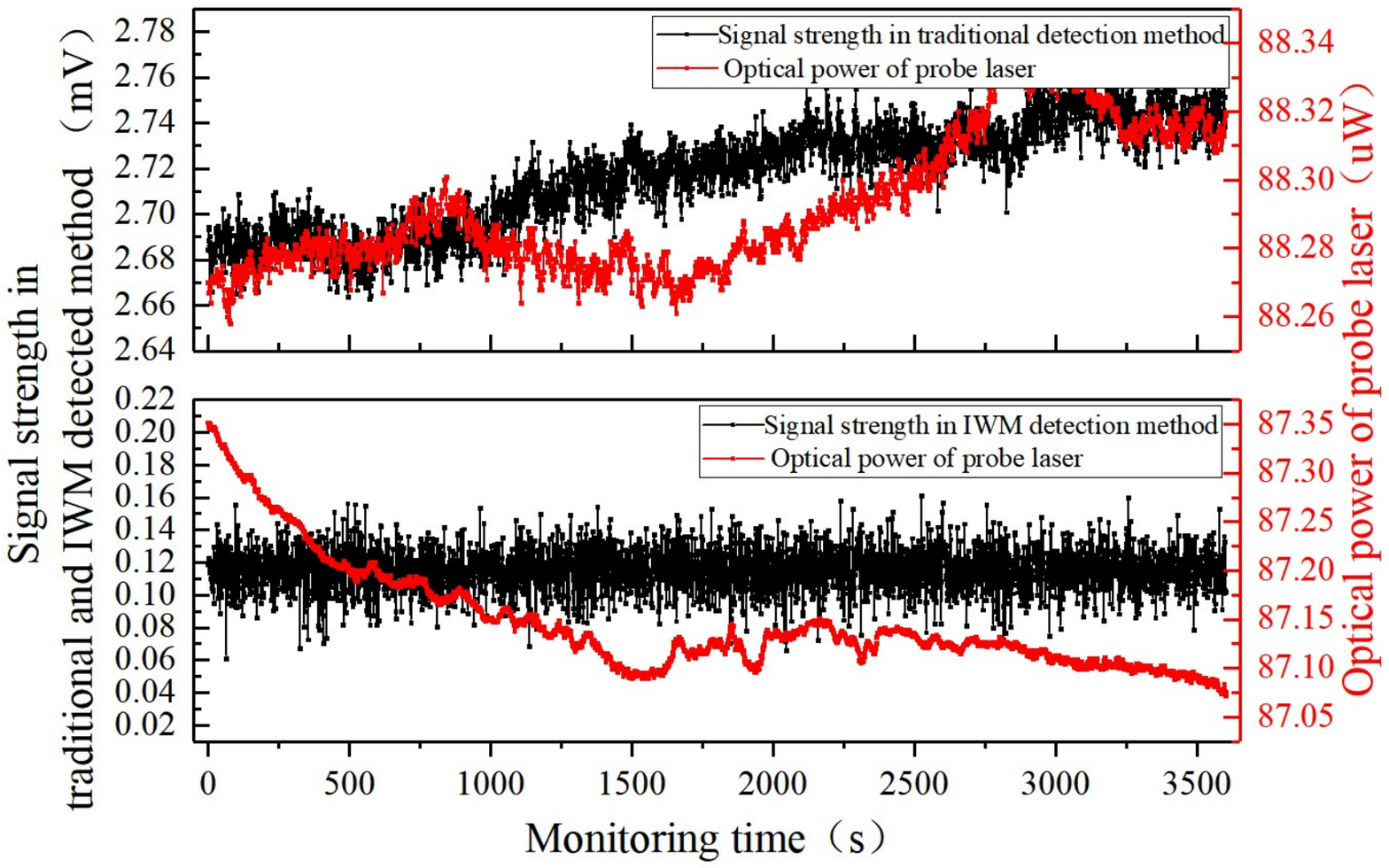}
\caption{Relationship between the optical power of the probe laser and the output signal in different detection methods. (a) traditional detection method, (b) IWM detection method.}
\label{Fig6}
\end{figure}

To assess the advantages of the IWM method in suppressing technical noise, the output signals from both the traditional detection method and the IWM method were continuously collected over a duration of one hour, while simultaneously monitoring and recording the optical power of the probe laser. As shown in Fig. \ref{Fig6} (a), the experimental results of the traditional detection method exhibit distinct correlations between the output signal and the optical power of the probe laser. This indicates that variations in the laser's optical power directly influence the output signal in the traditional method. In contrast, the IWM detection method, as described by Eqs. \eqref{eq7} and \eqref{eq8}, eliminates optical power-related terms from its output equation. This effectively prevents the output signal from being affected by fluctuations in the probe laser's optical power. While changes in optical power might still alter the overall brightness of the light spot, they do not result in any displacement of the spatial pattern in the detected signal. This key advantage of the IWM method is confirmed in Fig. \ref{Fig6}(b), where no correlation is observed between the optical power of the probe laser and the output signal.

\begin{figure}[ht]
\centering
\includegraphics[width=0.47\textwidth]{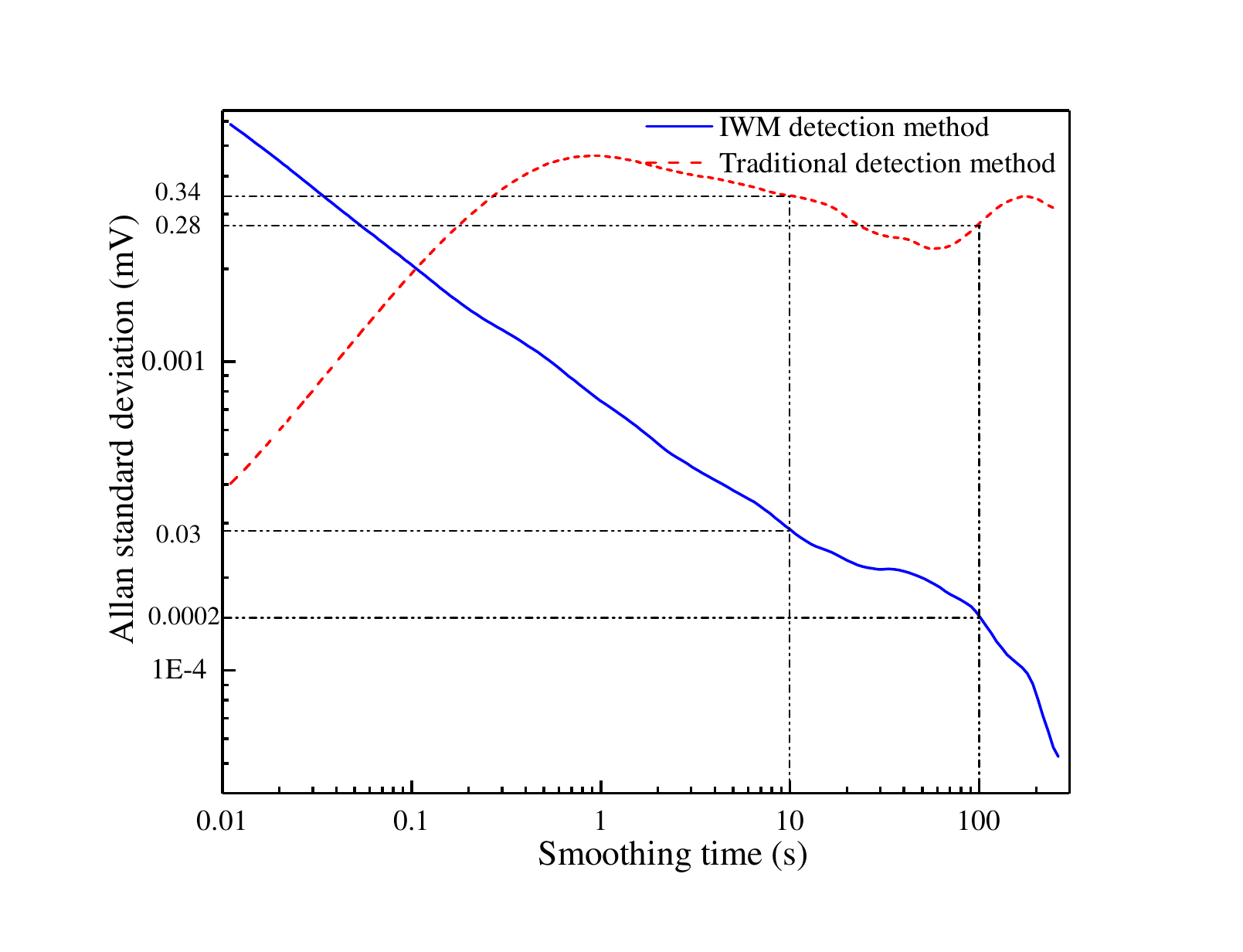}
\caption{Comparison of Allan standard deviations between the traditional detection method and the inverse weak measurement (IWM) detection method, with smoothing times ranging from 0.01 seconds to 1 hour.}
\label{Fig7}
\end{figure}

As a result, the IWM detection method significantly improves the stability of the output signal. To quantify this improvement, the Allan standard deviations of the two detection methods were calculated. As shown in Fig. \ref{Fig7}, the Allan standard deviations for the traditional and IWM detection methods are 0.34 and 0.28 for smoothing times of 10 seconds, and 0.03 and 0.002 for 100 seconds, respectively. Overall, the Allan standard deviation of the IWM detection method is consistently lower than that of the traditional method, indicating higher stability. By comparing the Allan standard deviations, the stability of the magnetic field output signal is improved by one to two orders of magnitude.

\section{IV. Conclusion}
In this work, we propose a novel detection method based on the inverse weak measurement (IWM), specifically designed for application in SERF magnetometers. In this approach, the magnetic field information is encoded into the polarization of the probe laser, while the transverse spatial pattern of the laser serves as the measurement pointer. The coupling between the quantum system (polarization) and the pointer enables precise extraction of the magnetic field strength through the displacement of the probe laser. Furthermore, post-selection of the polarization significantly enhances measurement sensitivity. The performance of the IWM-based detection method was rigorously evaluated in comparison with the traditional detection method. The IWM-enhanced SERF magnetometer achieves a magnetic field sensitivity of 182.8 $fT/Hz^{1/2}$, with results showing a one to two orders of magnitude improvement in stability, as reflected by the significantly lower Allan standard deviation. Our study not only validates the feasibility of incorporating weak measurement theory into SERF magnetometers but also highlights its potential to advance the performance of magnetic field measurement technology. These findings lay a solid foundation for further exploration and development of weak measurement-enhanced magnetometers.

	\textbf{Acknowledgments}
	The statistical support of this work was provided by Qian Cao, Jianqi Yang, Qianzi Yue. Writing assistance was provided by Qian Cao, Liang Xu. Optical Design support was provided by Qian Cao, Liang Xu, Yueyang Zhai. This work is funded by 2022 Industrial Technology Basic Public Service Platform Project under Grant (No. 2022189181), Innovation Program for Quantum Science and Technology (No. 2024ZD0300900), the National Natural Science Foundation of China (No. 12305034), the Natural Science Foundation of Jiangsu Province (No. BK20243060). The authors declare no conflicts of interest.

\end{document}